\documentclass[12pt]{article}
\usepackage[dvips]{graphicx}

\renewcommand{\bar}[1]{\overline{#1}}
\newcommand{\half}{{$\frac{1}{2}$}} 

\begin{document}

\begin{flushright}
SLAC-PUB-11747 \\
March, 2006
\end{flushright}



\def\bege{\begin{equation}}
\def\ende{\end{equation}}
\def\beger{\begin{eqnarray}}
\def\ender{\end{eqnarray}}

\def\vdp{{d^{4}p}\over{(2\pi)^{4}}}
\def\vdq{{d^{4}q}\over{(2\pi)^{4}}}
\def\half{{1\over 2}}
\def\slash#1{\;\raise.15ex\hbox{/}\kern-.77em #1}
\def\Dslash{\raise.15ex\hbox{/}\kern-.77em D}
\def\lsim{\mathrel{\mathstrut\smash{\ooalign{\raise2.5pt\hbox{$<$}\cr\lower2.5pt\hbox{$\sim$}}}}}

\def\ddp{d\tilde{p}}
\def\a{\alpha}
\def\b{\beta}
\def\D{\Delta}
\def\G{\Gamma}
\def\e{\epsilon}
\def\g{\gamma}
\def\d{\delta}
\def\k{\kappa}
\def\p{\phi}
\def\vp{\varphi}
\def\r{\rho}
\def\s{\sigma}
\def\l{\lambda}
\def\L{\Lambda}
\def\th{\theta}
\def\om{\omega}
\def\Om{\Omega}

\def\mub{\bar{\mu}}

\def\del{\partial}
\def\ha{\frac{1}{2}}
\def\bar#1{\overline{ #1 }}
\def\psibar{\overline{\psi}}
\def\etabar{\overline{\eta}}
\def\sla#1{#1\!\!\!/}
\def\smgroup{U(1)_Y{\otimes}SU(2)_L{\otimes}SU(3)_C}

\def\ra{\rangle}
\def\la{\langle}
\def\lraw{\leftrightarrow}
\def\raw{\rightarrow}
\def\bdlra{\buildrel \leftrightarrow \over}
\def\bdra{\buildrel \rightarrow \over}

\def\eg{{\it e.g.}}
\def\ie{{\it i.e.}}
\def\Dslash{\raise.15ex\hbox{/}\kern-.77em D}
\def\Aslash{\raise.15ex\hbox{/}\kern-.77em A}
\def\Hslash{\raise.15ex\hbox{/}\kern-.77em H}
\def\np#1#2#3{Nucl. Phys. {\bf #1} (#2) #3}
\def\pl#1#2#3{Phys. Lett. {\bf #1} (#2) #3}
\def\prl#1#2#3{Phys. Rev. Lett. {\bf #1} (#2) #3}
\def\pr#1#2#3{Phys. Rev. {\bf #1} (#2) #3}
\def\prd#1#2#3{Phys. Rev. D {\bf #1} (#2) #3}
\def\etal{{\it et al}}

\def\ddp{d\tilde{p}}
\def\ddl{d\tilde{l}}
\def\ddk{d\tilde{k}}
\def\g{\gamma}
\def\mm{m_{\pi}}
\def\ra{\rangle}
\def\la{\langle}
\def\gt{\tilde{\gamma}}
\def\mQbar{m_{\bar{Q}}}
\def\Lraw{\Longrightarrow}
\def\mgx{m_{\tilde{g}}}
\def\ggx{\tilde{g}}
\def\mqx{m_{\tilde{q}}}
\def\qx{\tilde{q}}
\def\mlx{m_{\tilde{l}}}
\def\lx{\tilde{l}}
\def\wx{\tilde{w}}
\def\tx{\tilde{t}}
\def\bx{\tilde{b}}
\def\hpm{{H}^{\pm}}
\def\hpmx{\tilde{H}^{\pm}}

\def\O{{\cal O}}
\def\M{{\cal M}}
\def\tilde{\widetilde}
\def\bar{\overline}
\def\Z{{\bf Z}}
\def\T{{\bf T}}
\def\S{{\bf S}}
\def\R{{\bf R}}
\def\np#1#2#3{Nucl. Phys. {\bf B#1} (#2) #3}
\def\pl#1#2#3{Phys. Lett. {\bf #1B} (#2) #3}
\def\prl#1#2#3{Phys. Rev. Lett.{\bf #1} (#2) #3}
\def\physrev#1#2#3{Phys. Rev. {\bf D#1} (#2) #3}
\def\ap#1#2#3{Ann. Phys. {\bf #1} (#2) #3}
\def\prep#1#2#3{Phys. Rep. {\bf #1} (#2) #3}
\def\rmp#1#2#3{Rev. Mod. Phys. {\bf #1} (#2) #3}
\def\cmp#1#2#3{Comm. Math. Phys. {\bf #1} (#2) #3}
\def\mpl#1#2#3{Mod. Phys. Lett. {\bf #1} (#2) #3}

\def\Lam#1{\Lambda_{#1}}
\def\pf{{\rm Pf ~}}
\font\zfont = cmss10 
\font\litfont = cmr6 \font\fvfont=cmr5
\def\bigone{\hbox{1\kern -.23em {\rm l}}}
\def\ZZ{\hbox{\zfont Z\kern-.4emZ}}
\def\half{{\litfont {1 \over 2}}}
\def\mx#1{m_{\hbox{\fvfont #1}}}
\def\gx#1{g_{\hbox{\fvfont #1}}}
\def\gG{{\cal G}}
\def\lamlam#1{\langle S_{#1}\rangle}
\def\CM{{\cal M}}
\def\CO{{\cal O}}
\def\Re{{\rm Re ~}}
\def\Im{{\rm Im ~}}
\def\lfm#1{\medskip\noindent\item{#1}}

\def\ku{\k_u}
\def\kup{\k_u'}
\def\kd{\k_d}
\def\kdp{\k_d'}
\def\Tr{{\rm Tr}}


\bigskip\bigskip

\begin{center}
{\Large \bf The Higgs Boson Mass \\
in Split Supersymmetry at Two-Loops\footnote{Work supported by the
Department of Energy under contract number DC-AC02-76SF00515}}
\end{center}

\vspace{13pt}

\centerline{ \bf Michael Binger}

\vspace{8pt} {\centerline{Stanford Linear Accelerator Center,}}

{\centerline{Stanford University, Stanford, California 94309,
USA}}

\centerline{e-mail: mwbinger@stanford.edu}

\bigskip\bigskip

\begin{abstract}

The mass of the Higgs boson in the Split Supersymmetric Standard
Model is calculated, including all one-loop threshold effects and
the renormalization group evolution of the Higgs quartic coupling
through two-loops. The two-loop corrections are very small ($\ll
1\;{\rm GeV}$), while the one-loop threshold corrections generally
push the Higgs mass down several ${\rm GeV}$.

\end{abstract}

\newpage

\section{Introduction}

The gauge hierarchy problem of the standard model (SM) of particle
physics has been a fruitful source of inspiration for beyond the SM
physics. Most notably, a main reason for the prominence of
supersymmetry was its natural solution to this problem. In recent
years, additional circumstantial evidence for supersymmetry (SUSY)
has arisen from gauge coupling unification and from dark matter,
although these successes have been partially offset by difficulties
with flavor changing neutral currents and CP violation which arise
from light SUSY scalars. Thus, it may be reasonable to abandon the
original motivation for SUSY and consider the implications of a
theory which maintains all of the successes of the MSSM, except for
the hierarchy problem, and does away with some of the difficulties.
This proposal, called finely tuned, or split supersymmetry, has
appeared in \cite{Arkani-Hamed:2004fb}\cite{Giudice:2004tc}, and
some phenomenology has been discussed
\cite{Mahbubani:2004xq,Mukhopadhyaya:2004cs,Pierce:2004mk,Arvanitaki:2004eu,Kilian:2004uj,Hewett:2004nw,Arkani-Hamed:2004yi}.
In split supersymmetry, a single Higgs scalar is fine tuned to be
light, with the understanding that the fine tuning will be resolved
by some anthropic-like selection effects. This approach may have a
natural realization within inflation and string theory
\cite{Susskind:2004uv}, where an almost infinite landscape of vacua
may contain a small percentage which have the desired fine-tuned
parameters necessary for life and the properties of our universe.

The prediction for the Higgs boson mass is typically higher in Split
SUSY than MSSM scenarios, and is thus a key distinguishing feature.
The MSSM Higgs mass is known to two-loop accuracy
\cite{MSSMhiggs2loop}. The purpose of this paper is to bring the
split supersymmetry Higgs mass prediction to a similar level of
precision.

\section{Corrections to the Higgs Mass}

The starting point for our analysis is the split SUSY Lagrangian
\cite{Arkani-Hamed:2004fb}\cite{Giudice:2004tc}
 \begin{eqnarray}
 {\cal L} &=& m^2 H^{\dagger}H - \frac{\l}{2}  (H^{\dagger}H)^2 +
 F_U\hat{H}^{\dagger} Q\bar{u}+ F_D H^{\dagger} Q\bar{d}
  + F_L H^{\dagger} L\bar{e} +{\rm h.c.}\nonumber\\
 &-&
 \frac{M_1}{2}\tilde{B}\tilde{B}-\frac{M_2}{2}\tilde{W}^I\tilde{W}^I
 -\frac{M_3}{2}\tilde{g}^a\tilde{g}^a - \mu \tilde{H}^T_u\e\tilde{H}_d\nonumber\\
 &-& \frac{H^{\dagger}}{\sqrt{2}}\Big( \ku\s^I\tilde{W}^I + \kup
 \tilde{B}  \Big)\tilde{H}_u +
 \frac{\hat{H}^{\dagger}}{\sqrt{2}}\Big( -\kd\s^I\tilde{W}^I +
 \kdp \tilde{B}  \Big)\tilde{H}_d +{\rm h.c.},
 \end{eqnarray}
where $\hat{H} = -i\s_2H^*$, $H=(H^+,H^0)^T$, and $\epsilon =
i\sigma_2$. The predictions for the Higgs mass will be derived from
this Lagrangian using methods similar to the work of Sirlin and
Zucchini on the SM Higgs boson \cite{Sirlin:1985ux} and the
subsequent work of Hempfling and Kniehl on the SM top quark
\cite{Hempfling:1994ar}.

After electroweak symmetry breaking, the bare Higgs mass is related
to the bare quartic coupling $\l_0$ and vacuum expectation value
(vev) $v_0 = \sqrt{2}\la H^0 \ra$ by
 \bege
 M_{h_0}(\mub) = \sqrt{\l_0(\mub)} v_0(\mub)
 \ende
where the dimensional regularization scale $\mub$ is introduced into
loop integrals through $\int d^4 x \raw \mub^{-2\e} \int d^d x$ and
is elevated to the renormalization scale in $\bar{MS}$. Each of
these bare quantities must be related to physical quantities in
order to obtain a meaningful relation.

The pole mass is related to the bare mass by
 \bege
  M_h^2 = M_{h_0}^2(\mub) + \Re\Sigma_h(M_h,\mub) + 3 \frac{T_h(\mub)}{v},
 \ende
where $\Sigma_h$ is the Higgs self energy and $T_h$ is the
tadpole\footnote{No tadpole counterterm is used in this paper. It is
a matter of convention whether or not one uses such a counterterm,
and the final results are easily seen to be independent of this
choice.}. The bare vev is related to the renormalized vev via muon
decay \cite{Sirlin:1980nh} :
 \bege
  v_0^2(\mub) = v_F^2\Bigg( 1+ \Pi_{WW}(0,\mub) + E(\mub) -2\frac{T_h(\mub)}{M_H^2 v_F}   \Bigg),
 \ende
where $v_F = 1/\sqrt{\sqrt{2}G_F} \; \approx \; 246.22 {\rm GeV}$,
$\Pi_{WW}(0,\mub)$ is the $W^{\pm}$ boson self-energy at zero
momentum, and $E$ represents vertex and box corrections to muon
decay in the standard model. Finally the bare coupling is related
to the $\bar{MS}$ coupling by
 \bege
 \l_0(\mub) = \l(\mub) + \frac{\b_\l}{2}C_{UV},
 \ende
where $C_{UV} = \frac{1}{\e} -\g_E +\log{4\pi} $ and $\b_\l =
\frac{\del\l}{\del\log{\mub}}$. Putting these together, one finds
 \begin{eqnarray}\label{mh}
 M_h &=& \sqrt{\l(\mub)}v_F\big(1+\d_h(\mub)\big)\nonumber\\
 \d_h(\mub)&=& \half\left( {\Re\Sigma_h(M_h,\mub)\over M_h^2} + \frac{T_h(\mub)}{M_h^2v_F} +
 E(\mub) + \Pi_{WW}(0,\mub) +\frac{\b_\l}{2\l}C_{UV}\right).
 \end{eqnarray}

This formula includes all one-loop threshold and renormalization
group (RG) corrections, and can be improved to include the two-loop
RG corrections to the running of $\lambda(\mub)$. The scale $\mub$
should be chosen to minimize large logarithmic corrections, although
at one-loop the $\mub$ (and $C_{UV}$) dependence formally cancels
from Eq.(\ref{mh}). The results for $\d_h(\mub)$ in the SM were
given in \cite{Sirlin:1985ux}\footnote{The convention used here is
related to \cite{Sirlin:1985ux} by $\d_h^{SM}({\rm this\;
paper})=-\half \d_h^{SM}(\cite{Sirlin:1985ux})$}. The split
supersymmetry threshold corrections are discussed in detail in
section \ref{thresh}.

\subsection{The algorithm used to calculate the Higgs
mass}\label{alg}

The input parameters for the Higgs mass analysis include
supersymmetry breaking scale $M_S$, $\tan{\beta}$ at $M_S$, and the
soft gaugino and higgsino masses $M_1$, $M_2$, $M_3$, and $\mu$ (not
to be confused with the RN scale $\mub$) which are specified at the
scale of gauge coupling unification $M_G\sim 3\times 10^{16}{\rm
GeV}$ and are assumed to be universal
 \bege
 M_{1/2} \equiv M_1(M_G) = M_2(M_G) = M_3(M_G) = \mu(M_G).
 \ende
Of course, the $\mu$-term may take different values from the
gaugino masses, but we have explicitly verified that the Higgs
mass prediction is very insensitive to the $\mu$ initial value, so
the results in Figs.(1-4) are valid for most other reasonable
values of $\mu$.

First, the coupled system of differential equations
\cite{Giudice:2004tc} for $g_1$, $g_2$, $g_3$, $F_U$, $F_D$, $F_L$,
$\ku$, $\kd$, $\kup$, $\kdp$ are solved numerically. The gauge
couplings are run at two loops, whereas the seven other couplings
are run at one loop. We keep only the top, bottom, and tau Yukawa
couplings and so can replace $F_U\rightarrow Y_t, F_D\rightarrow
Y_b, F_L\rightarrow Y_{\tau}$ in the all of the following formulae.
The boundary values of the gauge couplings $g_1,g_2,g_3$ and Yukawa
couplings $Y_b, Y_{\tau}$ are given at scale $M_Z$ from the latest
world averages \cite{Eidelman:2004wy}. As will be discussed in
section \ref{topyuk}, $Y_t$ is given at the top pole mass $M_t$,
including three-loop QCD corrections and one-loop threshold
corrections from electro-weak and split supersymmetric interactions.
The new split SUSY Yukawas are given at scale $M_S$ through the
relations:
 \begin{eqnarray}
 \ku(M_S) &=& g_2(M_S)\sin{\beta} \nonumber\\
 \kd(M_S) &=& g_2(M_S)\cos{\beta} \nonumber\\
 \kup(M_S) &=& \sqrt{\frac{3}{5}}g_1(M_S)\sin{\beta} \nonumber\\
 \kdp(M_S) &=& \sqrt{\frac{3}{5}}g_1(M_S)\cos{\beta}.
 \end{eqnarray}
Because the boundary values for the couplings are given at different
scales, it is necessary to take an iterative approach to solving the
differential equations. The couplings which are specified at low
scales, such as $y_t(M_t)$, are guessed at the high scale $M_s$, the
differential equations are then solved, and the resulting value for
$y_t(M_t)$ is compared to the correct value in order to obtain a
better guess, at which point the procedure is repeated. Five
iterations are usually sufficient. An additional complication arises
because the split SUSY corrections to $Y_t(M_t)$ (detailed in
section \ref{topyuk}) depend on $U,V,$ and $N$, which depend on the
solutions of the RGE's for the gaugino/higgsino masses, which depend
on the solutions of the RGE's for the dimensionless couplings, which
in turn depend upon $Y_t(M_t)$. Thus, this entire analysis should be
performed iteratively.

Armed with the RGE evolution of the dimensionless couplings, the
RGE's for the soft masses $M_1$, $M_2$, $M_3$, and $\mu$ are then
solved and are run down to scales $M_1$, $M_2$, $M_3$, and $\mu$,
respectively, where the physical pole masses are extracted, as
detailed in section \ref{spec}. The chargino and neutralino mixing
matrices $U,V,$ and $N$ will appear in the threshold corrections.

In section \ref{2loop} the two-loop running of the Higgs quartic
coupling will be given. The solutions for $g_1$, $g_2$, $g_3$,
$F_U$, $F_D$, $F_L$, $\ku$, $\kd$, $\kup$, $\kdp$ yield the required
inputs to solve the $\l$ RGE, with the gaugino and higgsino masses
providing the appropriate matching scale between the Standard Model
and split supersymmetric running. The boundary value of the Higgs
quartic coupling in minimal split supersymmetry is
 \bege
 \l(M_S) = \frac{1}{4} \Bigg( g_2^2(M_S)+\frac{3}{5}g_1^2(M_S) \Bigg)\cos^2{2\beta}.
 \ende

Finally, all that remains is to include the finite threshold
corrections in Eq.(\ref{mh}). These are detailed in section
\ref{thresh}. Results for the Higgs mass are given in section
\ref{results}.

\subsection{The gaugino and higgsino mass spectrum}\label{spec}
 The formulae for the running of the gaugino and higgsino masses
are given in Eqs.(57-63) of Ref.\cite{Giudice:2004tc}.

The gluino mass appears in our analysis as the threshold for the
running of the strong coupling. It is straightforward to evaluate
$M_3$ at scale $M_3$ and deduce the pole mass
 \bege
  M_{\tilde{g}} = M_3(M_3)\left( 1 + 12\;\frac{g_3^2(M_3)}{(4\pi)^2} \right).
 \ende

The chargino and neutralino mass matrix diagonalization proceeds
similar to the MSSM \cite{Haber:1984rc}. The mass matrices are
given by
 \bege
 X =  \Bigg( \matrix{ M_2 & \frac{\ku v}{\sqrt{2}} \cr \frac{\kd v}{\sqrt{2}}  & \mu } \Bigg)
\;\;\;\;\;\;\;\;\;\;
  Y=\left( \begin{array}{cccc} M_1 & 0 & -\frac{\kdp v}{\sqrt{2}} & \frac{\kup v}{\sqrt{2}}
 \\ 0 & M_2 & \frac{\kd v}{\sqrt{2}} & -\frac{\ku v}{\sqrt{2}}
 \\ -\frac{\kdp v}{\sqrt{2}}  & \frac{\kd v}{\sqrt{2}} & 0 & -\mu
 \\ \frac{\kup v}{\sqrt{2}} & -\frac{\ku v}{\sqrt{2}} & -\mu & 0 \end{array} \right)
 \ende

These are diagonalized by matrices $U,V$ in the chargino sector
($\chi_i^+ \chi_i^-$) and $N$ in the neutralino sector
($\chi_i^0$):
 \bege
  \chi_i^+ = V_{ij} \psi_j^+ \;\;\;\;\; \chi_i^- = U_{ij} \psi_j^- \;\;\;\;\chi_i^0 = N_{ij} \psi_j^0,
 \ende
where the gauge eigenstates are
 \bege \psi_j^+ = \Bigg( \matrix{ \tilde{W}^+ \cr \tilde{H}_u^+} \Bigg)
 \;\;\;\;\;\;
 \psi_j^- = \Bigg( \matrix{ \tilde{W}^- \cr \tilde{H}_d^-} \Bigg)
 \;\;\;\;\;\;
 \psi_j^0 = (\tilde{B}, \tilde{W}_3, \tilde{H}_d^0, \tilde{H}_u^0 )^T.
 \ende

The matrices $U,V,N$ are specified by
 \begin{eqnarray}\label{rotmat}
 N^* Y N^{-1} = M^{(N)} &=& {\rm diag}\{ M^{(N)}_1, M^{(N)}_2, M^{(N)}_3, M^{(N)}_4  \} \nonumber\\
 N Y^{\dagger} Y N^{-1} &=&  (M^{(N)})^2 \nonumber\\
 U^*XV^{-1} &=& M^{(C)} = {\rm diag}\{ M^{(C)}_1, M^{(C)}_2 \} \nonumber\\
 V X^{\dagger} X V^{-1} &=&  (M^{(C)})^2 = U^* X X^{\dagger} U^{T}
 \end{eqnarray}

The running mass parameters $M_1$, $M_2$, $\mu$ are evaluated at the
scales $M_1$, $M_2$, $\mu$, respectively, in an iterative fashion.
This minimizes the threshold corrections relating the pole masses
and running masses, which are not considered in detail. In any case,
the results are very insensitive to the exact scale chosen. The
threshold effects due to gaugino masses are incorporated into the
running of the dimensionless parameters appropriately.

\subsection{The Top Quark Yukawa coupling and pole
mass}\label{topyuk}

Since the Higgs mass is most sensitive to the top Yukawa coupling,
it is important to carefully extract this from the pole mass, which
is taken from the 2005 summer average of CDF and D0
\cite{cdftopmass} to be $M_t=172.7\pm 2.9 \;{\rm GeV}$. The leading
corrections are from QCD, and were calculated to two-loops in
\cite{Gray:1990yh} and to three-loops in \cite{chet3loopMQ}. The
full electro-weak corrections at one-loop were considered in
\cite{Hempfling:1994ar}, where the authors found the following
relation between the pole mass, the $\bar{MS}$ Yukawa coupling, and
the vacuum-expectation-value $v_F$ :
 \bege
  y_t(\mub) = \sqrt{2} \frac{M_t}{v_F} \Big(1+\d_t(\mub)\Big).
 \ende
The correction term is derived analogous to Eq.(\ref{mh}) and is
given by
 \bege
 \d_t(\mub) = {\rm Re}{\Sigma_t(M_t,\mub)} - \frac{\Pi_{WW}(0,\mub)}{2M_W^2} - \frac{E(\mub)}{2} - \frac{\b_{y_t}}{2y_t} C_{UV}.
 \ende
In this formula, $\Sigma_t$ represents the top quark self energy and
$E$ is the vertex and box corrections to muon decay, neither of
which receive new contributions in split SUSY at one-loop. However,
the $W$ boson self energy, $\Pi_{WW}$, does receive corrections,
which are calculated below. The UV divergence, $C_{UV}$, multiplying
the top Yukawa beta function $\b_{y_t}=\mub\frac{\del
y_t}{\del\mub}$ comes from the relation between the bare and
$\bar{MS}$ Yukawa coupling and is canceled by the divergent parts of
$\Sigma_t, \Pi_{WW},$ and $E$.

It is convenient to decompose the correction term into parts arising
from QCD, electro-weak theory (EW), and split-supersymmetry (SS) :
 \bege\label{dyt}
 \d_t(\mub) = \d_t^{QCD}(\mub)+\d_t^{EW}(\mub)+\d_t^{SS}(\mub)
 \ende
The three-loop QCD term derived from
\cite{Gray:1990yh}\cite{chet3loopMQ} for the top quark at scale
$\mub=M_t$ is
 \beger\label{dytqcd}
 \d_t^{QCD}(\mub=M_t) &\approx& -{4\over 3}\left({\a_3(M_t)\over \pi}\right)
 -9.1\;\left({\a_3(M_t)\over \pi}\right)^2-80\;\left({\a_3(M_t)\over \pi}\right)^3
 \nonumber\\
 &\approx& -0.046-0.011-0.003 \approx -0.060
 \ender
for $\a_3(M_Z)=0.118$. While the EW term given in
Ref.\cite{Hempfling:1994ar} formally depends on $M_h$ and $M_t$, it
turns out that in the range of Higgs and top masses of interest,
this contribution is negligible $|\d_t^{EW}|<0.001$.

Now we turn to the SS corrections, which arise only from the gaugino
and higgsino contribution to the $W^{\pm}$ self-energy. The
$W^{\pm}-\chi^{0}_i-\chi^{\pm}_j$  vertex is given by $ig
\g_{\mu}(L_{ij}P_L + R_{ij}P_R)$, with
 \beger
 L_{ij} &=& -\frac{1}{\sqrt{2}}N_{i4}V^*_{j2}+N_{i2}V^*_{j1}
 \nonumber\\
 R_{ij} &=& \frac{1}{\sqrt{2}}N^*_{i3}U_{j2}+N^*_{i2}U_{j1}.
 \ender
This vertex is used to derive the split SUSY corrections involving
charginos and neutralinos :
 \begin{eqnarray}\label{aww}
 16\!\!\!\!&\pi^2&\!\!\!\! \Pi_{WW}^{(C,N)}(0,\mub) = -2 M_W^2 C_{UV} X_2(SS)
 \\
 &+& g^2 \sum_{i=1}^{4} \sum_{j=1}^{2} \Bigg( (L_{ij}L_{ij}^*+R_{ij}R_{ij}^*)
 \left[ a^2 \left( \log{a^2\over \mub^2} - 1/2 \right)
 + b^2 \left( \log{b^2\over \mub^2} - 1/2 \right)
 +{a^2b^2\over a^2-b^2}\log{a^2\over b^2} \right]
 \nonumber\\
 &+& 2(L_{ij}R_{ij}^*+R_{ij}L_{ij}^*){ab\over a^2-b^2}
 \left[ -a^2\left(\log{a^2\over \mub^2} - 1\right) + b^2\left(\log{b^2\over \mub^2} - 1\right) \right]\Bigg),
 \end{eqnarray}
where we used the shorthand $a=M^{(C)}_j, b= M^{(N)}_i$ and
$X_2(SS)$ is given in Eq.(\ref{inv}). The resulting correction term
 \bege\label{dytss}
 \d_t^{SS}(\mub) = -{\Pi_{WW}^{(C,N)}(0,\mub)\over 2M_W^2}\Bigg|_{C_{UV}=0}
 \ende
depends on the soft gaugino mass terms, $\tan{\b}$, and the scalar
mass scale $M_S$. The scale $\mub$ must be chosen in accordance with
the decoupling scale $M_{SS}$ imposed on $y_t$ at the
chargino/neutralino thresholds, $\mub=M_{SS}$. The exact scale is
not very important, but consistently applying the choice to both the
running of $y_t$ and the threshold correction $\d_t^{SS}$ is
important. For the explicit results given in Figs.(1-4), the
decoupling scale was chosen to be the mass of the lightest
supersymmetric particle, which is typically a neutralino.
Generically the split SUSY correction is small,
$|\d_t^{SS}(\mub=M_{SS})| {\lsim} 0.01$, but should be included
since it can lead to a shift in the Higgs mass of up to $2\;{\rm
GeV}$.

To summarize, we have found
 \bege
 y_t(M_t) = 0.945\;\left({M_t\over 175\;{\rm GeV}}\right)(1+\d_t^{SS}(M_{SS}))
 \ende

\subsection{The 2-loop running of the Higgs Quartic
Coupling}\label{2loop}

It is useful to define the following invariants involving the
standard model Yukawas and the new split SUSY Yukawas :
 \begin{eqnarray}\label{inv}
 Y_2(SM) &=& {\rm Tr} \Big[3 F_U^{\dagger}F_U +3 F_D^{\dagger}F_D + F_L^{\dagger}F_L  \Big]  \nonumber\\
 Y_4(SM) &=& {\rm Tr} \Big[3 (F_U^{\dagger}F_U)^2 +3 (F_D^{\dagger}F_D)^2 + (F_L^{\dagger}F_L)^2  \Big]  \nonumber\\
 Y_6(SM) &=& {\rm Tr} \Big[3 (F_U^{\dagger}F_U)^3 +3 (F_D^{\dagger}F_D)^3 + (F_L^{\dagger}F_L)^3  \Big]  \nonumber\\
 Y_G(SM) &=& \left(\frac{17}{20}g_1^2+\frac{9}{4}g_2^2+8g_3^2 \right){\rm Tr} (F_U^{\dagger}F_U)\nonumber\\
 &+& \left(\frac{1}{4}g_1^2 +\frac{9}{4}g_2^2+8g_3^2\right){\rm Tr}
 (F_D^{\dagger}F_D) + \frac{3}{4}(g_1^2+g_2^2){\rm Tr}(F_L^{\dagger}F_L) \nonumber\\
 X_2(SS) &=& 3(\ku^2+\kd^2)+\kup^2+\kdp^2 \nonumber\\
 X_4(SS) &=& 5(\ku^4+\kd^4)+2\ku^2\kd^2+2(\ku\kup+\kd\kdp)^2+(\kup^2+\kdp^2)^2.
\end{eqnarray}

The running of the quartic coupling $\l$ of the Higgs boson is
governed by
 \bege
 \b_\l\;{\equiv}\;\mub\frac{\del\l}{\del\mub}
 =\frac{1}{16\pi^2}\b_\l^{(1)} +\frac{1}{(16\pi^2)^2}\b_\l^{(2)} +\cdots
 \ende
The one-loop beta function is given by \cite{Giudice:2004tc}
 \begin{eqnarray}
 \b_\l^{(1)} &=& 12\l^2 - 9\l \left( \frac{1}{5}g_1^2 + g_2^2 \right)
 +\left( \frac{27}{100} g_1^4 +\frac{9}{10}g_1^2g_2^2 +\frac{9}{4}g_2^4\right)
 \nonumber\\
 &+& 4\l Y_2(SM) -4Y_4(SM) + 2\l X_2(SS) - X_4(SS).
\end{eqnarray}

The 2 loop result is conveniently divided into two terms,
 \bege
 \b_\l^{(2)} = \b_\l^{(2)}(SM') + \b_\l^{(2)}(SS),
 \ende
where $SS$ is the new split SUSY contribution and $SM'$ denotes the
standard model result modified to include gauginos and higgsinos in
gauge boson self-energies. This is accomplished by replacing the
number of generations in the SM result with
 \bege
  N_g(1) = 3 + 3/10 =33/10 \;\; \;\;\;\;\;N_g(2) =3 + 3/2=9/2
 \ende
 \begin{eqnarray}\label{blam2SM}
 \b_\l^{(2)}(SM') &=& -78\l^3  -24 \l^2 Y_2(SM) -\l Y_4(SM) -42\l \Tr(F_U^{\dagger}F_U F_D^{\dagger} F_D) +20 Y_6(SM)
 \nonumber\\
 &-& 12 \Tr [ F_U^{\dagger}F_U (F_U^{\dagger}F_U + F_D^{\dagger} F_D ) F_D^{\dagger} F_D]
  +  10\l Y_G(SM) + 54 \l^2 \left(g_2^2 + {1\over 5}g_1^2 \right)
 \nonumber\\
 &-& \l \left[ \left( \frac{313}{8} -10N_g(2) \right)g_2^4 - \left( \frac{687}{200} + 2N_g(1) \right)g_1^4
  -\frac{117}{20}g_2^2g_1^2  \right]
 \nonumber\\
 &-& 64g_3^2\Tr\left[(F_U^{\dagger}F_U)^2 + (F_D^{\dagger} F_D)^2\right]
  - \frac{8}{5} g_1^2 \Tr\left[2(F_U^{\dagger}F_U)^2 - (F_D^{\dagger} F_D)^2 +3 (F_L^{\dagger}F_L)^2\right]
 \nonumber\\
 &-& \frac{3}{2}g_2^4 Y_2(SM) + g_1^2 \Bigg[ \left( \frac{63}{5}g_2^2-\frac{171}{50}g_1^2 \right)\Tr(F_U^{\dagger}F_U)
 + \left( \frac{27}{5}g_2^2+\frac{9}{10}g_1^2 \right)\Tr(F_D^{\dagger}F_D)
 \nonumber\\
 &+& \left( \frac{33}{5}g_2^2-\frac{9}{2}g_1^2 \right)\Tr(F_L^{\dagger}F_L) \Bigg]
 + \left( \frac{497}{8}-8N_g(2)\right) g_2^6 - \left( \frac{97}{40}+\frac{8}{5} N_g(2)  \right) g_2^4g_1^2
 \nonumber\\
 &-&\left( \frac{717}{200}+\frac{8}{5} N_g(1)  \right) g_2^2g_1^4 -
 \left( \frac{531}{1000}+\frac{24}{25} N_g(1)  \right) g_1^6
\end{eqnarray}

The new split SUSY Yukawas contribute
 \begin{eqnarray}\label{blam2SS}
 \b_\l^{(2)}(SS) &=& -12 \l^2 X_2(SS)  - \frac{\l}{4} \Big[ 5(\ku^4+\kd^4) + 44 \ku^2\kd^2
 + 2(\ku^2\kup^2+\kd^2\kdp^2) + \kup^4+\kdp^4
 \nonumber\\
 &-& 12\kup^2\kdp^2 - 80\ku\kd\kup\kdp \Big] + {47\over 2}(\ku^6+\kd^6)+{5\over 2}(\kup^6+\kdp^6)+{7\over 2}\ku^2\kd^2(\ku^2+\kd^2)
 \nonumber\\
 &+& {11\over 2}(\ku^4\kup^2+\kd^4\kdp^2) + {21\over 2}\ku^2\kd^2(\kup^2+\kdp^2) + 19\ku\kd\kup\kdp(\ku^2+\kd^2)
 \nonumber\\
 &+& 21\ku\kd\kup\kdp(\kup^2+\kdp^2) + {17\over 2}(\ku^2\kup^4+\kd^2\kdp^4+\kup^2\kdp^4+\kup^4\kdp^2)
 \nonumber\\
 &+&  {19\over 2}\kup^2\kdp^2(\ku^2+\kd^2)+ {15\over 4}\l\left[ (g_2^2+\frac{1}{5}g_1^2) X_2(SS)+8g_2^2(\ku^2+\kd^2) \right]
 \nonumber\\
 &-& 4g_2^2\left[ 5(\ku^4+\kd^4)+2\ku^2\kd^2+(\ku\kup+\kd\kdp)^2\right]-g_2^4\left[ \frac{3}{4}X_2(SS) +36(\ku^2+\kd^2)\right]
 \nonumber\\
 &+& \frac{3}{10}g_1^2g_2^2\left[ 21(\ku^2+\kd^2)-(\kup^2+\kdp^2)  \right]-\frac{9}{100}g_1^4X_2(SS).
 \end{eqnarray}

In deriving these, we relied on the useful papers of Luo, Wang, and
Xiao \cite{luo2loop}, which corrected some typos from the seminal
works of Machacek and Vaughn \cite{MV}. Also useful is
Ref.\cite{Arason:1991ic}.

\subsection{The Higgs Self Energy and Tadpole
Corrections}\label{thresh}

In split supersymmetry there are corrections to $\Sigma_h$, $T_h$,
and $\Pi_{WW}$, but not to $E$.

The Higgs tadpole and self-energy depend on the mass mixing matrices
which appear in the Feynman rules. The interaction Lagrangian in
terms of the physical mass eigenstate Dirac and Majorana fermions is
given by
 \bege
 {\cal L}_{int} = -\frac{h}{\sqrt{2}}\bar{\psi}_i^{\pm}(P_L L^C_{ij}+P_R R^C_{ij})\psi^{\pm}_j
 +\frac{h}{2}\bar{\psi}_i^{0} (P_L (R^N_{(ij)})^* + P_R R^N_{(ij)})\psi^{0}_j,
 \ende
where the mixing matrices are
 \begin{eqnarray}
 R^C_{ij}&=&(L^C_{ji})^*= \ku V_{i2} U_{j1}+\kd V_{i1}U_{j2} \nonumber\\
 R^N_{ij}&=& (\ku N_{i2}-\kup N_{i1}) N_{j4} - (\kd N_{i2} - \kdp N_{i1}) N_{j3} \nonumber\\
 R^N_{(ij)} &=& \frac{1}{2}(R^N_{ij} + R^N_{ji}).
 \end{eqnarray}

The split-supersymmetric contribution to the Higgs tadpole $iT_h
=i(T_h^{(C)}+T_h^{(N)})$ involves charginos and neutralinos :
 \bege
 16\pi^2 T_h^{(C)}(\mub) = -2\sqrt{2}\sum_{i=1}^{2}\Re
 \left[ R^C_{ii}(M^C_i)^3\left( C_{UV}-\log{\frac{(M^C_i)^2}{\mub^2}}+1\right)\right]
 \ende
 \bege
 16\pi^2 T_h^{(N)}(\mub) = 2\sum_{i=1}^{4}\Re\left[ R^N_{(ii)}(M^N_i)^3\left(C_{UV}-\log{\frac{(M^C_i)^2}{\mub^2}}+1\right) \right].
 \ende

The Higgs self energies are easily written in terms of the canonical
one-loop basis functions $A(M)$, $B_0(k^2;M_1,M_2)$
\cite{Bohm:1986rj}:
 \beger
 16\pi^2 \Sigma_h^{(C)}(p^2,\mub) \!&=&\! \sum_{i,j=1}^{2} \Bigg[ \half (|L^C_{ij}|^2+|R^C_{ij}|^2)
 \Big( A(M_i)+ A(M_j) +(M_i^2+M_j^2-p^2)B_0(p^2;M_i,M_j)  \Big)
 \nonumber\\
 &+&\! 2\Re M_iM_j R^C_{ij}(L^C_{ij})^* B_0(p^2;M_i,M_j) \Bigg]
 \nonumber\\
 16\pi^2 \Sigma_h^{(N)}(p^2) &=& \sum_{i,j=1}^{4} \Bigg[ |R^N_{(ij)}|^2
 \Big( A(M_i)+ A(M_j) +(M_i^2+M_j^2-p^2)B_0(p^2;M_i,M_j)  \Big)
 \nonumber\\
 &+& 2\Re M_iM_jR^N_{(ij)}(R^N_{(ij)})^* B_0(p^2;M_i,M_j) \Bigg]
\end{eqnarray}
The above results lead to the following correction term for the
Higgs mass:
 \bege\label{dhss}
 \d_h^{SS}(\mub) = \half\left( {\Re{\left(\Sigma_h^{(C)}(M_h,\mub)+\Sigma_h^{(N)}(M_h,\mub)\right)} \over M_h^2}
  + \frac{T_h^{(C)}(\mub)+T_h^{(N)}(\mub)}{M_h^2v_F} + \Pi_{WW}^{(CN)}(0,\mub) \right)\Bigg|_{C_{UV}=0}.
 \ende
As discussed below Eq.(\ref{dytss}), the scale $\mub$ must be chosen
in accordance with the decoupling imposed on $\l$ at the
chargino/neutralino thresholds, $\mub=M_{SS}$. Combined with the SM
results of \cite{Sirlin:1985ux}, which should be evaluated at $M_h$,
the total threshold correction is given by
 \beger\label{dhtot}
 M_h &=& \sqrt{\l(M_h)}v_F\big(1+\d_h(M_h)\big)
 \nonumber\\
 \d_h(M_h) &=& \d_h^{SM}(M_h) + \d_h^{SS}(M_{SS}).
 \ender
A useful check of these results is the cancelation of divergences
$C_{UV}$ in Eq.(\ref{mh}). This involves repeated use of the
definitions in Eq.(\ref{rotmat}), and has been explicitly verified.

\section{Results for the Higgs Mass}\label{results}

The corrections to the Higgs mass considered in this paper are of
three varieties:
\begin{itemize}
\item{{\bf Top Yukawa Coupling.} The threshold corrections to the
Yukawa coupling initial value given in Eq.(\ref{dyt}) are amplified
because $y_t$ is raised to the fourth power in $\b_\l^{(1)}$. The
QCD corrections to $y_t(M_t)$ are dominant ($\sim -6\%$) and lead to
a downward shift in the Higgs mass of about $15\;{\rm GeV}$. The
electro-weak corrections are negligible over the entire parameter
range of interest.

The split SUSY correction is small but not negligible. For each
choice of parameters $\tan{\b}$, $M_s$, and $M_{1/2}$, there will be
a correction term $\d_t^{SS}$ to the initial value $y_t(M_t)$.
However, $y_t(M_t)$ is required input for solving the coupled
differential equations which eventually lead to $\d_t^{SS}$. Thus,
in principle an iterative approach must be taken. After performing
this type of analysis we found that it could be circumvented by
using the output $\d_t^{SS}$ along with the following simple rule of
thumb : every shift in $y_t(M_t)$ of $\pm 0.0045$ will shift the
Higgs mass by $\pm 1\;{\rm GeV}$. The contributions of the bottom
and $\tau$ Yukawa couplings turn out to be completely negligible and
can be omitted from the beginning.}

\item{{\bf Two-loop running of $\l$.} The two-loop correction to
the beta function is numerically very small, which is partially due
to cancelations between the SM and split SUSY contributions in
Eqs.(\ref{blam2SM},\ref{blam2SS}). The shift in the Higgs mass due
to including $\b_\l^{(2)}$ is less than $300\;{\rm MeV}$ for all
relevant values of $M_s$, $\tan{\b}$, and $M_{1/2}$.}

\item{{\bf Threshold corrections ($\d_h$).} The correction given in
Eq.(\ref{dhtot}) typically pushes down the Higgs mass by several
${\rm GeV}$, with a larger shift occurring for small $\tan\b$ and
small $M_s$. Typically, the SM contributes most of this shift, with
the split SUSY corrections $\lsim 1\;{\rm GeV}$.}
\end{itemize}

All of the above corrections should be considered in the context
of two sources of uncertainty. First, the uncertainties in the top
mass $M_t = 172.7{\pm}2.9\;{\rm GeV}$ and $\a_s(M_z)=0.118\pm
0.003$ translate into uncertainties in $M_h$ of about $\pm (3-5)\;
{\rm GeV}$ and $ \mp (0.3-1.2) \;{\rm GeV}$, respectively. Second,
there are model specific ``theory uncertainties" at the high scale
\cite{Mahbubani:2004xq}.

Some representative plots of the Higgs mass are shown in Figs.(1,2).
In Figs.(3,4), the two-loop and threshold corrections discussed
above are plotted. The large QCD corrections to $M_h$
($\sim-15\;{\rm GeV}$) arising from Eq.(\ref{dytqcd}) are not shown
explicitly in Figs.(3,4) in order to clearly illustrate the other
much smaller effects.

\begin{figure}[h]
 \centering \includegraphics[angle=270,width=16cm]{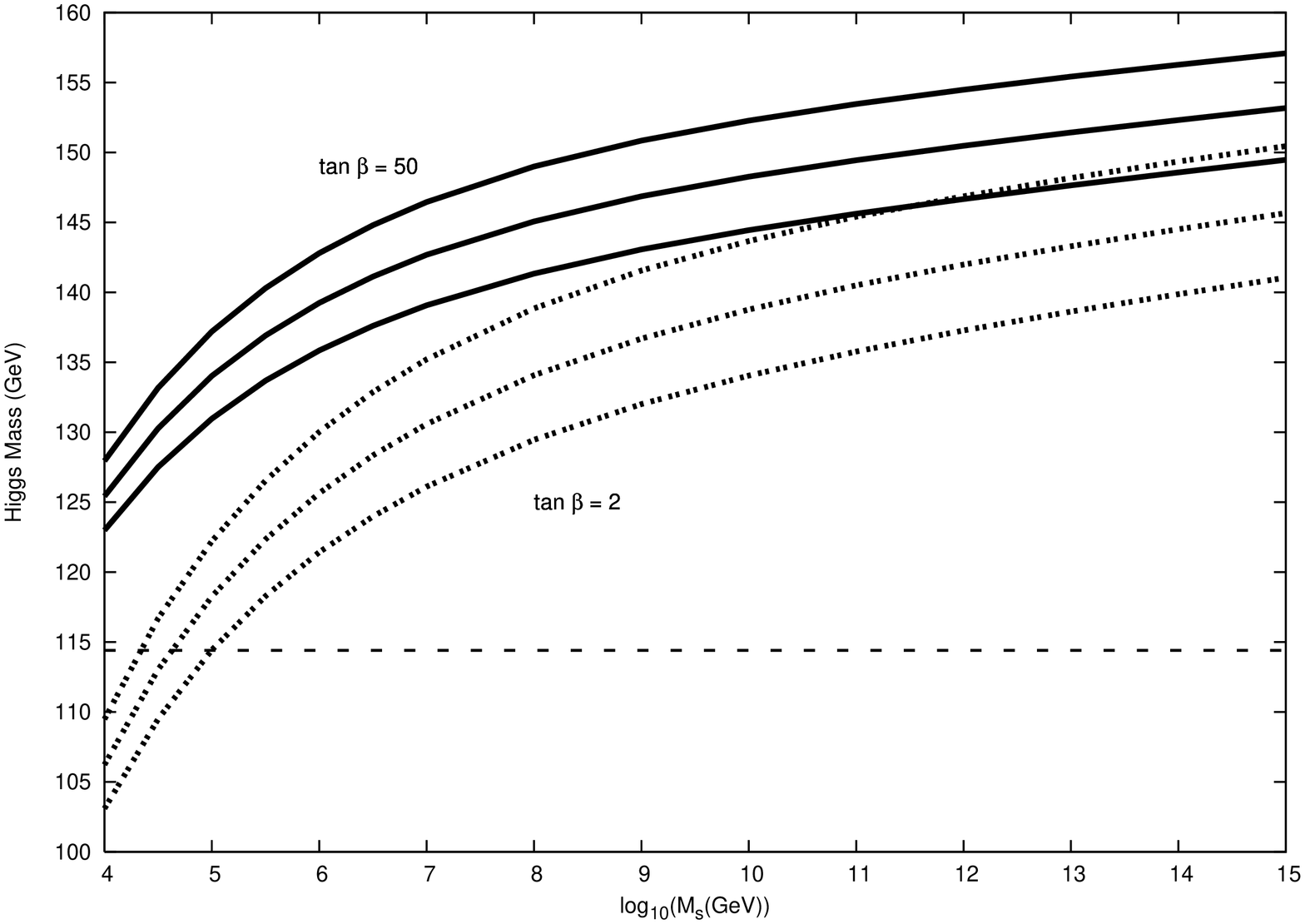}
 \caption{ The Higgs mass prediction versus $M_s$ for $\tan{\b}=2$ (lower dotted lines)
 and $\tan{\b}=50$ (upper solid lines). For each set of three the middle
 line is with $\a_s(M_z)=0.118$, $M_t=172.7$; the upper line is with
 $\a_s(M_z)=0.115$, $M_t=175.6$; and the lower line is with
 $\a_s(M_z)=0.121$, $M_t=169.8$. These correspond to the $1\sigma$
 variations of $\a_s(M_Z)=0.118\pm 0.003$ and $M_t=172.7\pm 2.9 \;{\rm GeV}$,
 with the resulting uncertainties in the Higgs mass considered additively.
 The gaugino and higgsino masses at $M_G$ are taken as universal
 $M_{1/2}=500\;{\rm GeV}$. The experimental lower bound
 \cite{Barate:2003sz} of $M_h>114.4\;{\rm GeV}$ (at 95\%) is
 shown.
 }\label{fig1}
\end{figure}

\begin{figure}[h]
 \centering \includegraphics[angle=270,width=16cm]{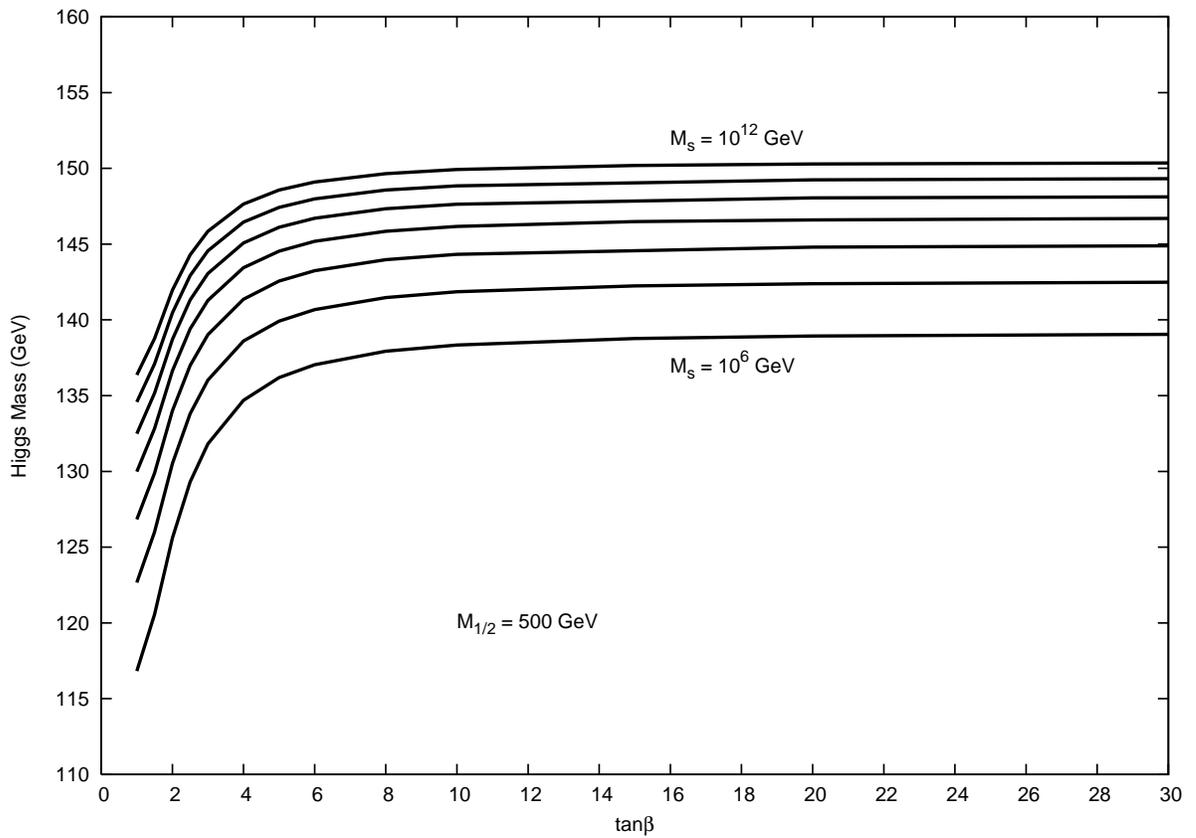}
 \caption{The Higgs mass versus $\tan{\b}$, for
 $M_s=10^6,10^7,10^8,10^9,10^{10},10^{11}$, and $10^{12}\;{\rm GeV}$, from bottom to top.
 Here $\a_s(M_z)=0.118$ and $M_t=172.7$.}\label{fig2}
\end{figure}

\begin{figure}[h]
 \centering \includegraphics[angle=270,width=16cm]{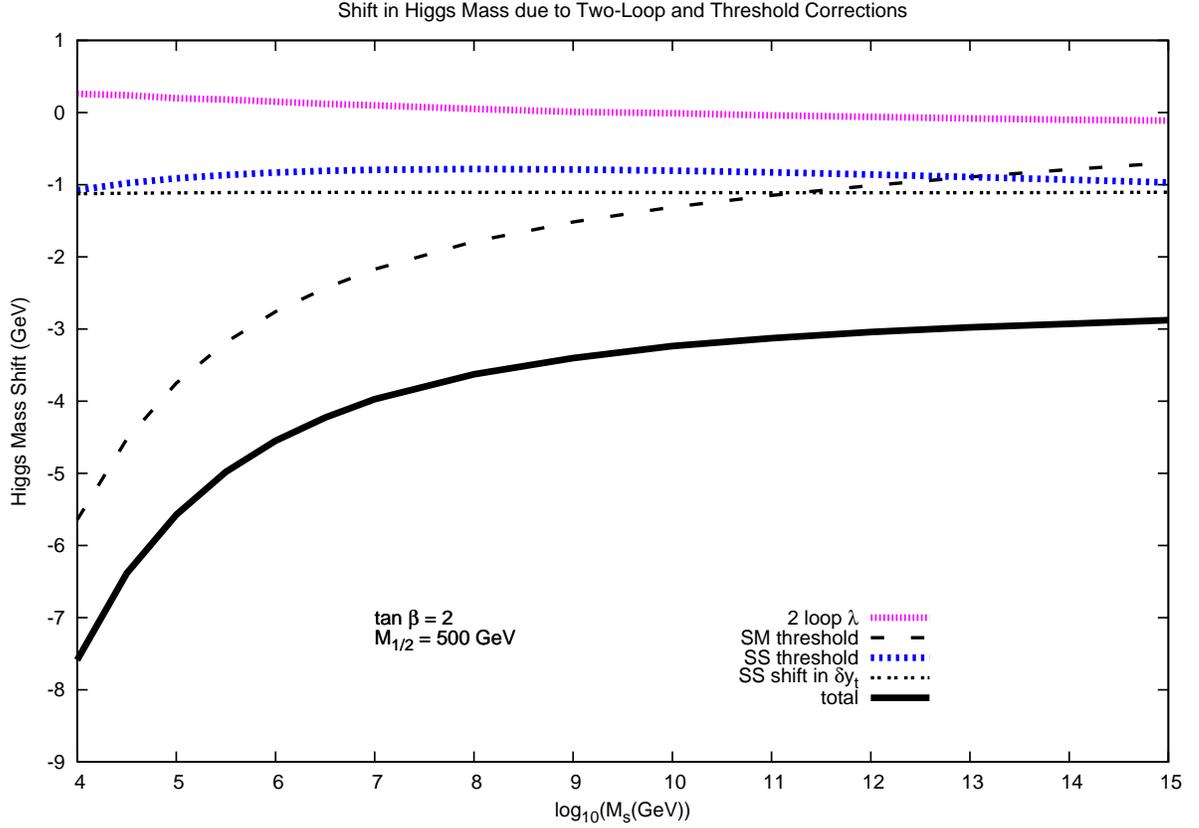}
 \caption{ The Higgs mass shift for $\tan{\b}=2$ and $M_{1/2}=500\;{\rm GeV}$
 due to four types of threshold corrections. The correction due to the two-loop
 running of $\l$ is shown in the thick dense dotted line near zero. The
 Standard Model (SM) correction from \cite{Sirlin:1985ux} is
 the dashed line. The split SUSY (SS) correction from Eq.(\ref{dhss})
 is the thick dotted line. The thin dotted line is the SS correction to the Higgs
 mass through the correction to the top Yukawa initial value,
 Eq.(\ref{dytss}). The solid line is the total of these four
 corrections. Here $\a_s(M_z)=0.118$ and $M_t=172.7$.
  }\label{fig3}
\end{figure}

\begin{figure}[h]
 \centering \includegraphics[angle=270,width=16cm]{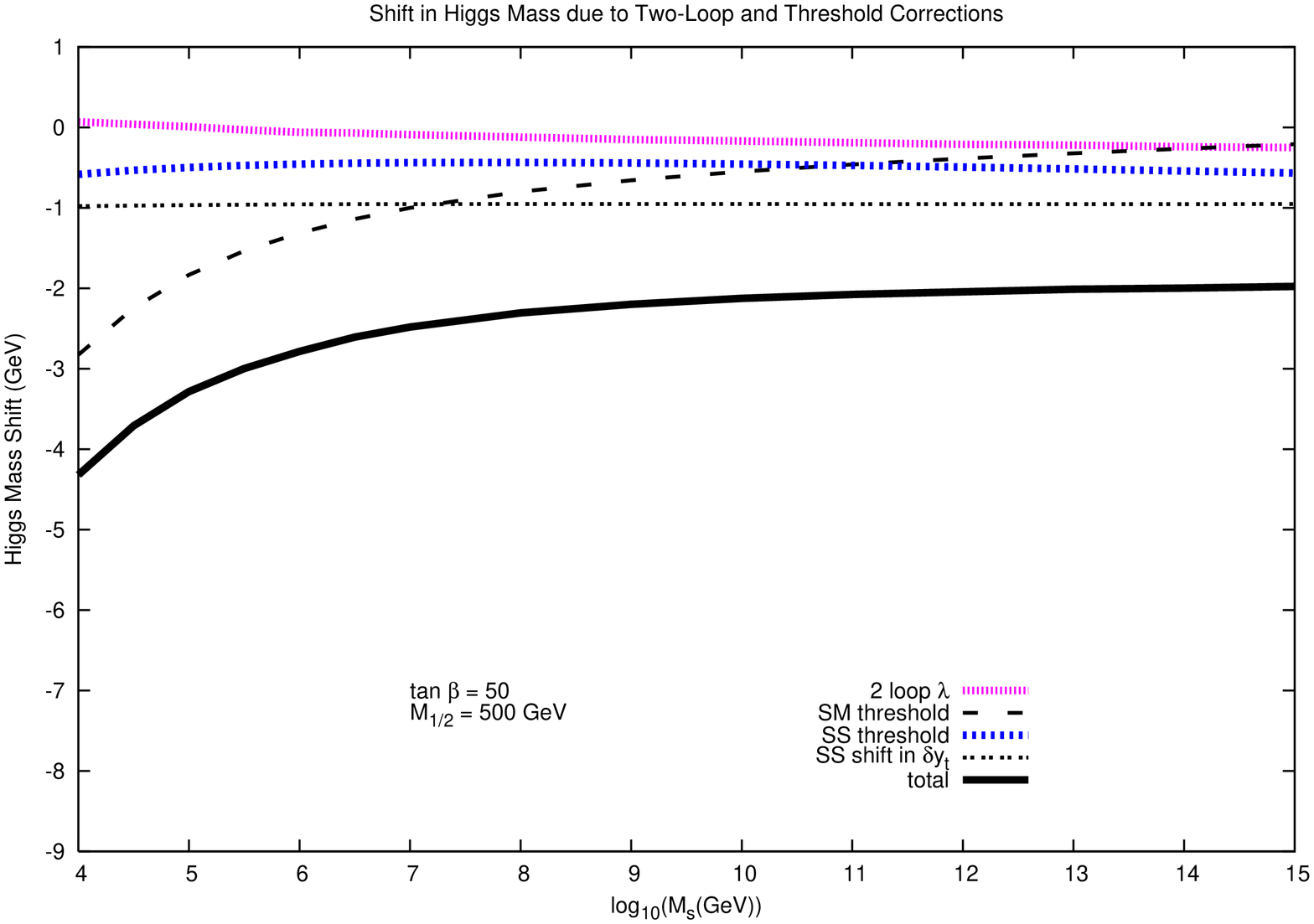}
 \caption{Same as Fig.3 except with $\tan{\b} =50$.}\label{fig4}
\end{figure}

\newpage



\begin{thebibliography}{99}

\bibitem{Arkani-Hamed:2004fb}
  N.~Arkani-Hamed and S.~Dimopoulos,
  JHEP {\bf 0506}, 073 (2005)
  [arXiv:hep-th/0405159].

\bibitem{Giudice:2004tc}
  G.~F.~Giudice and A.~Romanino,
  Nucl.\ Phys.\ B {\bf 699}, 65 (2004)
  [Erratum-ibid.\ B {\bf 706}, 65 (2005)]
  [arXiv:hep-ph/0406088].

\bibitem{Mahbubani:2004xq}
  R.~Mahbubani,
  arXiv:hep-ph/0408096.

\bibitem{Mukhopadhyaya:2004cs}
  B.~Mukhopadhyaya and S.~SenGupta,
  Phys.\ Rev.\ D {\bf 71}, 035004 (2005)
  [arXiv:hep-th/0407225].

\bibitem{Pierce:2004mk}
  A.~Pierce,
  Phys.\ Rev.\ D {\bf 70}, 075006 (2004)
  [arXiv:hep-ph/0406144].

\bibitem{Arvanitaki:2004eu}
  A.~Arvanitaki, C.~Davis, P.~W.~Graham and J.~G.~Wacker,
  Phys.\ Rev.\ D {\bf 70}, 117703 (2004)
  [arXiv:hep-ph/0406034].

\bibitem{Kilian:2004uj}
  W.~Kilian, T.~Plehn, P.~Richardson and E.~Schmidt,
  Eur.\ Phys.\ J.\ C {\bf 39}, 229 (2005)
  [arXiv:hep-ph/0408088].

\bibitem{Hewett:2004nw}
  J.~L.~Hewett, B.~Lillie, M.~Masip and T.~G.~Rizzo,
  JHEP {\bf 0409}, 070 (2004)
  [arXiv:hep-ph/0408248].

\bibitem{Arkani-Hamed:2004yi}
  N.~Arkani-Hamed, S.~Dimopoulos, G.~F.~Giudice and A.~Romanino,
  Nucl.\ Phys.\ B {\bf 709}, 3 (2005)
  [arXiv:hep-ph/0409232].

\bibitem{Susskind:2004uv}
  L.~Susskind,
  arXiv:hep-th/0405189.

\bibitem{MSSMhiggs2loop}
  R.~Hempfling and A.~H.~Hoang,
  Phys.\ Lett.\ B {\bf 331}, 99 (1994)
  [arXiv:hep-ph/9401219].
  H.~E.~Haber, R.~Hempfling and A.~H.~Hoang,
  Z.\ Phys.\ C {\bf 75}, 539 (1997)
  [arXiv:hep-ph/9609331].
  M.~Carena, M.~Quiros and C.~E.~M.~Wagner,
  Nucl.\ Phys.\ B {\bf 461}, 407 (1996)
  [arXiv:hep-ph/9508343].

\bibitem{Sirlin:1985ux}
  A.~Sirlin and R.~Zucchini,
  Nucl.\ Phys.\ B {\bf 266}, 389 (1986).

\bibitem{Hempfling:1994ar}
  R.~Hempfling and B.~A.~Kniehl,
  Phys.\ Rev.\ D {\bf 51}, 1386 (1995)
  [arXiv:hep-ph/9408313].

\bibitem{Sirlin:1980nh}
  A.~Sirlin,
  Phys.\ Rev.\ D {\bf 22}, 971 (1980).

\bibitem{Eidelman:2004wy}
  S.~Eidelman {\it et al.}  [Particle Data Group],
  Phys.\ Lett.\ B {\bf 592}, 1 (2004).

\bibitem{Haber:1984rc}
  H.~E.~Haber and G.~L.~Kane,
  Phys.\ Rept.\  {\bf 117}, 75 (1985).

\bibitem{cdftopmass}
    [CDF Collaboration],
  arXiv:hep-ex/0507091.

\bibitem{Gray:1990yh}
  N.~Gray, D.~J.~Broadhurst, W.~Grafe and K.~Schilcher,
  Z.\ Phys.\ C {\bf 48}, 673 (1990).

\bibitem{chet3loopMQ}
  K.~G.~Chetyrkin and M.~Steinhauser,
  Phys.\ Rev.\ Lett.\  {\bf 83}, 4001 (1999)
  [arXiv:hep-ph/9907509].
  Nucl.\ Phys.\ B {\bf 573}, 617 (2000)
  [arXiv:hep-ph/9911434].

\bibitem{luo2loop}
  M.~x.~Luo and Y.~Xiao,
  Phys.\ Rev.\ Lett.\  {\bf 90}, 011601 (2003)
  [arXiv:hep-ph/0207271].
  M.~x.~Luo, H.~w.~Wang and Y.~Xiao,
  Phys.\ Rev.\ D {\bf 67}, 065019 (2003)
  [arXiv:hep-ph/0211440].

\bibitem{MV}
  M.~E.~Machacek and M.~T.~Vaughn,
  Nucl.\ Phys.\ B {\bf 222}, 83 (1983).
  Nucl.\ Phys.\ B {\bf 236}, 221 (1984).
  Nucl.\ Phys.\ B {\bf 249}, 70 (1985).

\bibitem{Arason:1991ic}
  H.~Arason, D.~J.~Castano, B.~Keszthelyi, S.~Mikaelian, E.~J.~Piard, P.~Ramond and B.~D.~Wright,
  Phys.\ Rev.\ D {\bf 46}, 3945 (1992).


\bibitem{Bohm:1986rj}
  M.~Bohm, H.~Spiesberger and W.~Hollik,
  Fortsch.\ Phys.\  {\bf 34}, 687 (1986).

\bibitem{Barate:2003sz}
  R.~Barate {\it et al.}  [LEP Working Group for Higgs boson searches],
  Phys.\ Lett.\ B {\bf 565}, 61 (2003)
  [arXiv:hep-ex/0306033].



\end{thebibliography}
\end{document}